\documentclass{article}
 
\usepackage{latexsym}
\usepackage{ejs-article}

\usepackage[english]{babel}
\usepackage{graphicx}

\begin{document}

\Issue{8}{1}{35-47}{2008}

\HeadingAuthor{C. Sarraute and J. Burroni}
\HeadingTitle{OS Fingerprinting}

\Title{Using Neural Networks to improve classical Operating System Fingerprinting techniques} 

\Author {Carlos Sarraute \inst{1,2} \And Javier Burroni \inst{3} }

\institute{ Doctorado en Ingeniería Informática del ITBA
(Instituto Tecnológico de Buenos Aires). 
Av. Eduardo Madero 399, Buenos Aires, Argentina.
\and Corelabs research team. Core Security Technologies. \\
{\tt carlos@coresecurity.com}
\and Core Impact development team. Core Security Technologies. \\
{\tt jburroni@coresecurity.com}
}
\institutename

\begin{abstract}
We present remote Operating System detection as an inference problem: 
given a set of observations (the target host responses to a set of tests), 
we want to infer the OS type which most probably generated these observations.
Classical techniques used to perform this analysis present several limitations. 
To improve the analysis, we have developed tools using neural networks and Statistics tools. 
We present two working modules: 
one which uses DCE-RPC endpoints to distinguish Windows versions, 
and another which uses Nmap signatures to distinguish different version of 
Windows, Linux, Solaris, OpenBSD, FreeBSD and NetBSD systems. 
We explain the details of the topology and inner workings of the neural networks used, 
and the fine tuning of their parameters. 
Finally we show positive experimental results.
\end{abstract}
\smallskip

{\bf Keywords.} {\small Neural Networks, OS Fingerprinting, DCE-RPC endpoint mapper.}

\Body

\section{Introduction}

We present an application of Artificial Intelligence techniques to the field of Information Security.
The problem of remote Operating System (OS) Detection, also called
OS Fingerprinting, is a crucial step of the penetration testing process,
since the attacker (hacker or security professional)
needs to know the OS of the target host in order to choose the exploits that 
he will use. OS Detection is accomplished by passively sniffing network packets
and actively sending test packets to the target host, to study specific variations in
the host responses revealing information about its operating system.

The first fingerprinting implementations were based on the analysis of differences
between TCP/IP stack implementations.  The next generation focused the
analysis on application layer data such as the DCE RPC endpoint
information. 
Even though more information was analyzed,
some variation of the ``best fit" algorithm was still used to
interpret this new information.  
This strategy presents some weaknesses:
it will not work in non-standard situations and is unable to extract
the key elements that uniquely identify an operating system. 
We think that the next step is
to focus on the algorithm used to analyze the data rather than the data
itself.

Our new approach involves an analysis of the composition of the
information collected during the OS identification process to identify key elements and their
relations. 
To implement this approach, we have developed tools using Neural Networks
and techniques from the field of Statistics.
These tools have been successfully integrated in a commercial software (Core Impact).

\section{DCE-RPC Endpoint mapper}

\subsection{Information given by the DCE-RPC service}

In Windows systems, the Distributed Computing Environment (DCE) Remote Procedure Call (RPC) 
service allows connections bound for a target host's port 135. 
By sending an RPC query, you can determine which services or programs are registered with 
the RPC endpoint mapper database on the destination computer. 

The response includes the universal unique identifier (UUID) for each program, 
the annotated name (if one exists), the protocol that each program uses, 
the network address that the program is bound to, and the program's endpoint.
It's possible to distinguish Windows versions, editions and service packs based on the
combination of endpoints provided by the DCE-RPC service.

For example, in a Windows 2000 Professional edition service pack 0 machine, 
the RPC service returns 8 endpoints corresponding to 3 programs:

\begin{verbatim}
uuid="5A7B91F8-FF00-11D0-A9B2-00C04FB6E6FC"
annotation="Messenger Service"
 protocol="ncalrpc"      endpoint="ntsvcs"       id="msgsvc.1" 
 protocol="ncacn_np"     endpoint="\PIPE\ntsvcs" id="msgsvc.2" 
 protocol="ncacn_np"     endpoint="\PIPE\scerpc" id="msgsvc.3" 
 protocol="ncadg_ip_udp"                         id="msgsvc.4" 

uuid="1FF70682-0A51-30E8-076D-740BE8CEE98B"
 protocol="ncalrpc"      endpoint="LRPC"         id="mstask.1" 
 protocol="ncacn_ip_tcp"                         id="mstask.2" 

uuid="378E52B0-C0A9-11CF-822D-00AA0051E40F"
 protocol="ncalrpc"      endpoint="LRPC"         id="mstask.3" 
 protocol="ncacn_ip_tcp"                         id="mstask.4" 
\end{verbatim}

\subsection{Neural Networks come into play}

Our idea is to model the function which maps endpoints combinations to
operating system versions with a neural network.
We chose to use a multilayer perceptron network,
more precisely composed of 3 layers
(we indicate the number of neurons in each layer between parentheses).
\begin{enumerate}
\item{The input layer (with 413 neurons)
 contains one neuron for each UUID
 and one neuron for each endpoint corresponding to that UUID.
 Following the previous example, we have one neuron for the Messenger service
 and 4 neurons for each endpoint associated to that program.
 This enables us to respond with flexibility to the appearance of an unknown endpoint:
we still retain the information of the main UUID.}
\item{The hidden neurons layer (with 42 neurons),
 where each neuron represents a combination of the input neurons.}
\item{The output layer (with 25 neurons),
 contains a neuron for each version and edition of Windows
(e.g. Windows 2000 professional edition),
and one neuron for each version and service pack of Windows
 (e.g. Windows 2000 service pack 2). 
This way the network can distinguish the edition and the service pack independently:
errors in one dimension do not affect errors in the other dimension.}
\end{enumerate}


Network training is done using the backpropagation algorithm.
Given an expected output, an estimation of the error $\delta_{i,j}$ is calculated and 
propagated to the previous layers.
The weights, at each iteration $t$, are updated by adding $\Delta w_t$ 
that depends on a correction factor
and also on the value of
$\Delta w_{t-1}$ 
multiplied by a momentum $\mu$ 
(this gives the modifications a kind of kinetic energy):
$$
\Delta w_{t; i,j,k} = ( \lambda \cdot \delta_{i+1,k} \cdot v_{i,j}  ) + \mu \cdot \Delta w_{t-1; i,j,k} 
$$
The correction factor depends on the calculated $\delta$ values
and also on a learning rate $\lambda$ 
which can be adjusted to speed up training convergence.

The type of training realized is supervised training 
(based on a dataset containing inputs and expected outputs).
One generation consists in recalculating the synaptic weights
for each input / output pair.
At the beginning of each generation,
the inputs are reordered randomly,
so that the order of samples in the dataset doesn't affect training.

Complete training requires 10350 generations,
which can take 14 hours of python code execution.
Given that the design of the network topology is a trial-and-error
process, that requires training the network for each
variation of the topology to see if it produces better results,
the execution time motivated us to improve the training convergence speed 
(problem that we will tackle in section 4).

\subsection{Results}

The following table shows a comparison (from our laboratory) between
the old DCE-RPC module (which uses a ``best fit'' algorithm)
and the new module which uses a neural network to analyze the information.

\begin{center}
\begin{tabular} {| l | c | c | }
\hline
Result & Old DCE-RPC & DCE-RPC with \\
 & module & neural networks \\
\hline
Perfect match & 6 & 7 \\
Partial match & 8 & 14 \\
Mismatch & 7 & 0 \\
No answer & 2 & 2 \\
\hline
\end{tabular}
\end{center}

We reproduce below the result of executing the module against 
a Windows 2000 server edition SP1 machine. 
The correct system is recognized with precision.

\begin{verbatim}
Neural Network Output (close to 1 is better):
Windows NT4: 4.87480503763e-005
Editions:
    Enterprise Server: 0.00972694324639
    Server: -0.00963500026763
Service Packs:
    6: 0.00559659167371
    6a: -0.00846224120952
Windows 2000: 0.996048928128
Editions:
    Server: 0.977780526016
    Professional: 0.00868998746624
    Advanced Server: -0.00564873813703
Service Packs:
    4: -0.00505441088081
    2: -0.00285674134367
    3: -0.0093665583402
    0: -0.00320117552666
    1: 0.921351036343
Windows 2003: 0.00302898647853
Editions:
    Web Edition: 0.00128127138728
    Enterprise Edition: 0.00771786077082
    Standard Edition: -0.0077145024893
Service Packs:
    0: 0.000853988551952
Windows XP: 0.00605168045887
Editions:
    Professional: 0.00115635710749
    Home: 0.000408057333416
Service Packs:
    2: -0.00160404945542
    0: 0.00216065240615
    1: 0.000759109188052
Setting OS to Windows 2000 Server sp1
Setting architecture: i386
\end{verbatim}

\section{OS Detection based on Nmap signatures}

\subsection{Wealth and weakness of Nmap}

Nmap is a network exploration tool and security scanner which includes an OS detection method 
based on the response of the target host to a set of 9 tests. 
Below is a brief description of 
the packets sent in each test, for more information see \cite{fyodor}.

\begin{center}
\begin{tabular}{| l | l | l | l  |}
\hline
Test  & send packet   &  to port  &  with flags enabled \\
\hline
T1  &  TCP  &  open TCP  &  SYN, ECN-Echo  \\
T2  &  TCP  &  open TCP  &  no flags \\
T3  &  TCP  &  open TCP  &  URG, PSH, SYN, FIN \\
T4  &  TCP  &  open TCP  &  ACK \\
T5  &  TCP  &  closed TCP  &  SYN \\
T6  &  TCP  &  closed TCP  &  ACK \\
T7  &  TCP  &  closed TCP  &  URG, PSH, FIN \\
PU &  UDP  &  closed UDP  &  \\
TSeq &  TCP * 6  & open TCP  & SYN \\
\hline
\end{tabular}
\end{center}

Our method is based on the Nmap signature database. A signature is a set of rules
describing how a specific version / edition of different operating systems
responds to these tests. For example:

\begin{verbatim}
# Linux 2.6.0-test5 x86
Fingerprint Linux 2.6.0-test5 x86
Class Linux | Linux | 2.6.X | general purpose
TSeq(Class=RI%gcd=<6%SI=<2D3CFA0&>73C6B%IPID=Z%TS=1000HZ)
T1(DF=Y%W=16A0%ACK=S++%Flags=AS%Ops=MNNTNW)
T2(Resp=Y%DF=Y%W=0%ACK=S%Flags=AR%Ops=)
T3(Resp=Y%DF=Y%W=16A0%ACK=S++%Flags=AS%Ops=MNNTNW)
T4(DF=Y%W=0%ACK=O%Flags=R%Ops=)
T5(DF=Y%W=0%ACK=S++%Flags=AR%Ops=)
T6(DF=Y%W=0%ACK=O%Flags=R%Ops=)
T7(DF=Y%W=0%ACK=S++%Flags=AR%Ops=)
PU(DF=N%TOS=C0%IPLEN=164%RIPTL=148%RID=E%RIPCK=E%UCK=E%ULEN=134%DAT=E)
\end{verbatim}

The Nmap database contains 1684 signatures, which means some 1684 different
operating systems versions / editions could theoretically be distinguished by this method.

Given a host response to the tests, Nmap works by comparing it to each signature in the
database. A score is assigned to every signature, which is simply
the number of matching rules divided by the number of considered rules (the signatures may have
different number of rules, or some responses may be missing in which case the rule is not taken
into account). 
Thus Nmap performs a ``best score" or ``best fit" algorithm based on a Hamming distance between the response and the signatures,
where all the fields have the same weight.

This method presents the following problem:
improbable operating systems, which generate less responses to the
tests, get a better score (the matching rules acquire more relative weight). 
It happened during our tests that Nmap detected 
an OpenBSD 3.1 as as ``Foundry FastIron Edge Switch (load balancer) 2402",
or a Linux Mandrake 7.2 box as a ``ZyXel Prestige Broadband router."
The wealth of the database becomes a weakness!

\subsection{Hierarchical Network Structure}

If we represent the operating system space symbolically 
as a space in 568 dimensions 
(we will explain later the reason of this number),
the possible responses of the different system versions included in the database
form a cloud of points.
This large cloud has a particular structure, given that the OS families form
clusters which can be more or less recognized.
The Nmap method consists in finding the closest point to a given host response
(using the mentioned Hamming distance).

Our approach consists in several steps: 
in first place, we filter the operating systems that are not relevant
(according to the attacker's point of view, that is the operating systems for 
which he doesn't have exploits).
In our implementation, we are interested in the families
Windows, Linux, Solaris, OpenBSD, NetBSD and FreeBSD.
Then we assign the machine to one of the 6 considered families, 
therefore making use of the underlying family structure.
The result is a module which uses several neural networks, 
organized as a hierarchical network structure:
\begin{enumerate}
\item{
First step, a neural network to decide if the OS is relevant or not.}
\item{
Second step, a neural network to decide the OS family:
Windows, Linux, Solaris, OpenBSD, FreeBSD, NetBSD.}
\item{
In the Windows case, we use the
DCE-RPC endpoint mapper module to refine detection.}
\item{
In the Linux case, we realize a conditioned analysis to distinguish kernel version.}
\item{
In the Solaris and BSD cases, we realize a conditioned analysis to distinguish version.}
\end{enumerate}

We use a different neural network for each analysis, so we have
5 neural networks, and each requires a special topology and training.

\subsection{Neural network inputs}

The first question to solve is how to translate the host responses to neural network inputs.
We assign a set of input neurons for each test. 
The details for the test T1 $\ldots$ T7 are:\\
$\cdot \,$ One neuron for the ACK flag.\\
$\cdot \,$ One neuron for each response: S, S++, O.\\
$\cdot \,$ One neuron for the DF flag.\\
$\cdot \,$ One neuron for the response (yes/no).\\
$\cdot \,$ One neuron for the Flags field.\\
$\cdot \,$ One neuron for each response flag: ECN-Echo, URG,
ACK, PSH, RST, SYN, FIN. (in total 8 neurons).\\
$\cdot \,$ 10 groups of 6 neurons for the Options field. We activate one neuron
in each group, according to the option in the successive string positions: 
EOL, MAXSEG, NOP, TIMESTAMP, WINDOW, ECHOED (in total 60 neurons).  \\
$\cdot \,$ One neuron for the W field, which takes as input an hexadecimal value (the window size).

For the neurons corresponding to flags or options, the input is 1 or -1 (present or absent).
Other neurons have a numerical input, such as the field W (window size), the field GCD (the greatest common divisor
of the initial sequence numbers), or the SI and VAL responses to the TSeq test.
In the example of the Linux 2.6.0 response:
\begin{verbatim}
T3(Resp=Y%DF=Y%W=16A0%ACK=S++%Flags=AS%Ops=MNNTNW)
\end{verbatim}
maps to:\\
\begin{tabular} {| l | l | l | l | l |  l | l | l | l | l |  l | l | l | l | l |    }
\hline
ACK & S & S++ & O & DF & Yes & Flags & E & U & A & P & R & S & F & $\ldots$ \\
\hline
1     & -1 & 1     & -1 &  1  &   1   & 1        & -1    &  -1     &   1    &   -1    &  -1   &   1   &  -1   &  $\ldots$ \\
\hline
\end{tabular}

That way we obtain an input of 568 dimensions, with a certain amount of redundancy. 
This redundancy gives flexibility to our method when faced to unknown responses, 
but also raises performance issues!
We will see in the next section how to deal with this problem (reducing the number of dimensions).
As with the DCE-RPC module, the neural network consist of three layers. 
For example, the first neural network (relevance filter) contains:
the input layer 96 neurons, the hidden layer 20 neurons, the output layer 1 neuron.

\subsection{Dataset generation}

To train the neural network, we need a fair amount of inputs (host responses) with 
their corresponding outputs (host OS).
As the signature database contains 1684 rules, we estimated that a population of
15000 machines would be needed to train the network. As we don't have access
to a population of that size,  and scanning the Internet is not an option!

The adopted solution is to generate inputs by Monte Carlo simulation.
For each rule, we generate a number of inputs matching that rule, 
the actual number depending on the empirical distribution of operating systems.
When the rule specifies a constant value, we use that value, and when the rule
specifies options or a range of values, we chose a value following a uniform distribution
between the range of values.

\section{Dimension reduction and training}

\subsection{Correlation matrix}

In the design of the network topology, we have been generous with the input,
allowing 568 dimensions (with important redundancy).
This causes the training convergence to be slow, 
specially taking into account the size of the inputs dataset.
Our solution to this issue was to reduce the number of input dimensions.
This analysis also gives us insight about the key elements of the Nmap tests.

We consider each input dimension as a random variable $X_i (1 \leq i \leq 568)$.
As the input dimensions have different orders of magnitude (the flags take 0/1 values, others
are integers, like the ISN - initial sequence number), we first normalize the
random variables.
Then we compute the correlation matrix $R$ given by
$$
R_{i,j} = \frac{ E[ (X_i - \mu_i ) (X_j - \mu_j) ] } { \sigma_i \; \sigma_j }
$$
Since after normalization, $\mu_i = 0$ and $\sigma_i = 1$ for all $i$, 
the correlation matrix is simply $ R_{i,j} = E[ X_i \,X_j ] $.
The correlation is a measure of statistical dependence between 
two variables (closer to 1 or -1 indicates higher dependence).
Thus linear dependent columns of $R$ indicate linear dependent variables,
and we can safely keep one and eliminate the others, since they don't provide
additional information.
This analysis also results in the elimination of constants, which have zero variance.

Let's look at the results in the case of OpenBSD systems.
We reproduce below an extract of the signatures of 2 different OpenBSD systems,
where the fields that survive the correlation matrix reduction are marked in italics.

\noindent \texttt{Fingerprint OpenBSD 3.6 (i386) \\
Class OpenBSD | OpenBSD | 3.X | general purpose \\
T1(DF=N \% \emph{W=4000} \% ACK=S++ \% Flags=AS \% Ops=\emph{MN}WNNT) \\
T2(Resp=N) \\
T3(\emph{Resp=N}) \\
T4(DF=N \% \emph{W=0} \% ACK=O \%Flags=R \% Ops=)   \\
T5(DF=N \% W=0 \% ACK=S++ \% Flags=AR \% Ops=)  }

\noindent \texttt{Fingerprint OpenBSD 2.2 - 2.3 \\
Class OpenBSD | OpenBSD | 2.X | general purpose \\
T1(DF=N \% \emph{W=402E} \% ACK=S++ \% Flags=AS \% Ops=\emph{MN}WNNT) \\
T2(Resp=N) \\
T3(\emph{Resp=Y} \% DF=N \% \emph{W=402E} \% ACK=S++ \% Flags=AS \% Ops=\emph{MN}WNNT) \\
T4(DF=N \% \emph{W=4000} \% ACK=O \% Flags=R \% Ops=) \\
T5(DF=N \% W=0 \% ACK=S++ \% Flags=AR \% Ops=)
}

For example, in the test T1, the only fields that present variations
are $W$ and the first two options,
the others are constants in all the OpenBSD versions.
Another example, in the test T4 only $W$ may present variations,
and the test T5 doesn't give any information on the version of the examined OpenBSD system.

After performing this analysis, only 34 fields remain (out of 568).
These are the fields that
give information about the different OpenBSD versions.
As we have said, the test T5 doesn't appear, 
whereas tests Tseq and PU maintain several variables,
this shows us that these two tests are the most discriminative ones
between the OpenBSD population.

\subsection{Principal Component Analysis}

Further reduction is performed by means of Principal Component Analysis (PCA).
The idea of PCA is to compute a new basis (coordinate system) of the input space,
such that the greatest variance of any projection of the dataset 
comes by projecting to the first $k$ basis vectors. The parameter $k$ was chosen so as to keep 98\%
of the total variance.

After performing PCA we obtain the following topologies for the neural networks
(the size of the original input layer is 568 in all cases):

\noindent
\begin{center}
\begin{tabular} { | l | c | c | c | c | }
\hline
Analysis &  Input layer & Input layer & Hidden  & Output  \\
 & (after reduction & (after PCA) & layer & layer \\
& of matrix $R$ & & & \\
\hline
Relevance  & 204 & 96 & 20 & 1 \\
OS family   & 145 & 66 & 20 & 6 \\
Linux          & 100 & 41 & 18 & 8 \\
Solaris       & 55 & 26 & 7 & 5 \\
OpenBSD & 34 & 23 & 4 & 3 \\
\hline
\end{tabular}
\end{center}

To conclude the OpenBSD example, from the 34 variables that 
survived the correlation matrix reduction,
it's possible to construct a new basis with 23 vectors.
The coordinates in that base are the inputs for the neural network,
the hidden layer only contains 4 neurons and the output layer 3 neurons
(because we only distinguish 3 groups of OpenBSD versions).
Once we know that a machine is an OpenBSD,
the problem of recognizing the version is much simpler and bounded,
and can be accomplished by a smaller neural network (faster and more efficient).

\subsection{Adaptive learning rate}

This a strategy to accelerate the training convergence.
The learning rate is the parameter $\lambda$ that appears
in the back propagation formulas.

Given a network output, we can calculate an estimation of the quadratic error
$$
\frac { \sum_{i=1}^{n} ( y_i - v_i ) ^ 2 } {n}
$$
where $y_i$  are the expected outputs and $v_i$ are the network outputs.

After each generation (that is after processing all the input / output pairs),
if the error increases, we diminish the learning rate.
On the contrary, is the error gets smaller, we increase the learning rate.
The idea is to move faster if we are in the correct direction.

\begin{figure}
\centering
\includegraphics[width=11cm]{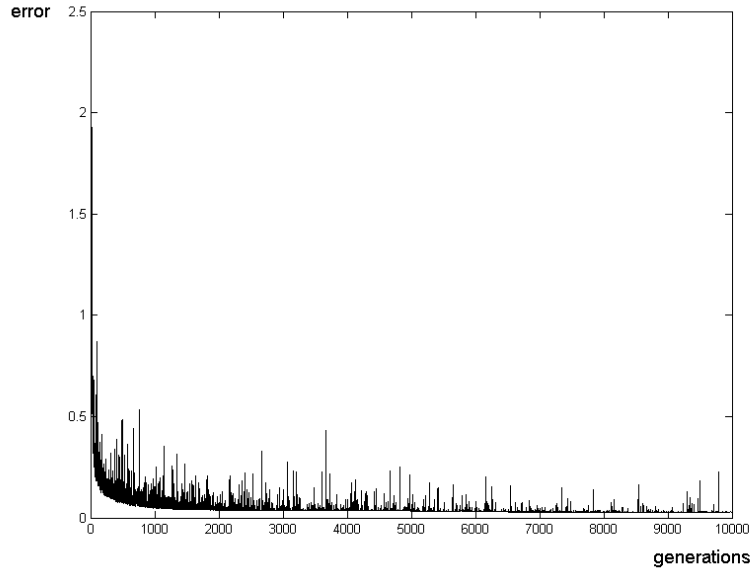}
\caption{Fixed learning rate} \label{fig-fixed}
\end{figure}

\begin{figure}
\centering
\includegraphics[width=11cm]{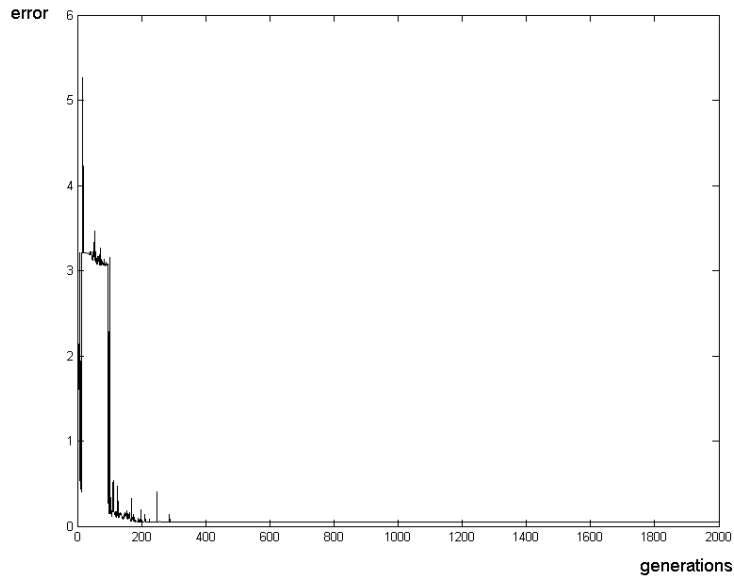}
\caption{Adaptive learning rate} \label{fig-adaptive}
\end{figure}

The figures \ref{fig-fixed} and \ref{fig-adaptive} show the evolution of the 
mean quadratic error as a function of the number of generations, for each strategy.
With a fixed learning rate, the error decreases and reaches satisfying levels after 4000 or 5000 generations.
By using an adaptive learning rate, we obtain in the first generations a more chaotic behavior,
with higher error levels.
But once the system finds the correct direction, the error drops rapidly to reach a low and constant value
after 400 generations.
These results are clearly better and provide a faster training of the network.

\subsection{Sample results}

We reproduce below the result of executing the Neural Nmap module
against a host running Solaris 8.
The correct system is recognized with precision.
\begin{verbatim}
Relevant / not relevant analysis
    0.99999999999999789    relevant 

Operating System analysis
    -0.99999999999999434   Linux 
    0.99999999921394744    Solaris 
    -0.99999999999998057   OpenBSD
    -0.99999964651426454 	 FreeBSD 
    -1.0000000000000000    NetBSD
    -1.0000000000000000    Windows

Solaris version analysis
    0.98172780325074482    Solaris 8 
    -0.99281382458335776   Solaris 9 
    -0.99357586906143880   Solaris 7 
    -0.99988378968003799   Solaris 2.X 
    -0.99999999977837983   Solaris 2.5.X 
\end{verbatim}

And below are the results of a comparison between the Neural Nmap module 
and the ``classic" Nmap module (that uses the scoring algorithm).
In the case of Windows systems, the respective DCE-RPC modules are used to refine
the version and edition detection.
The OS detection using neural networks is finer (more {\em version and edition} matches and
less {\em mismatches}).

\begin{center}
\begin{tabular} {| l | c | c | }
\hline
Result & Classic module & Neural Networks module \\
\hline
Version and edition match & 3 & 24 \\
Version match & 12 & 14 \\
Partial match &  13 & 7 \\
Only family match & 12  &  1 \\
Mismatch & 9 & 2 \\
No answer & 0 & 1 \\
\hline
\end{tabular}
\end{center}

\section{Conclusion and ideas for future work}

One of the main limitations of classical OS Detection techniques is the analysis performed on
the data collected by the tests, based on some variation of the ``best fit''
algorithm (compute the closest point according to a Hamming distance).

We have described how to generate and collect the information to be analyzed and
how to extract the structure of the input data.
This is the main idea of our approach, which motivates the decision to use neural networks,
to divide the analysis in several hierarchical steps,
and to reduce to input dimensions.
The experimental results (from our laboratory) show that this approach gives
a more reliable OS recognition.

Besides, the reduction of the correlation matrix and the principal component analysis
give us a systematic method to analyze the response of a host to the stimuli that we send.
As a result, we can identify the key elements of the Nmap tests,
for example the fields that give information about the different OpenBSD versions.
A further application of this analysis would be to optimize the Nmap tests to generate less traffic.
Another more ambitious application would be to create a database with responses of a
representative population to a large battery of tests
(combinations of different types of packets, ports and flags).
The same methods of analysis would allow to find in this vast database the most discriminative
tests for OS recognition.

The analysis that we propose can also be applied to other detection methods:
\begin{enumerate}
\item{Xprobe2, by Ofir Arkin, Fyodor Yarochkin \& Meder Kydyraliev, 
that bases the detection on ICMP, SMB and SNMP tests.}

\item{Passive OS Identification (p0f) by Michal Zalewski,
method that has the advantage of generating zero additional traffic.
It is an interesting challenge, since the analysis involves an important volume of information
(all the sniffed traffic), and probably requires more dynamic and evolutive methods.}

\item{OS Detection based on SUN RPC portmapper information,
that allows distinction of Sun, Linux and other versions of System V.}

\item{Information gathering for client-side penetration test,
in particular to detect versions of applications.
For example to distinguish Mail User Agents (MUA) such as Outlook or Thunderbird,
by extracting information of the mail headers.}

\end{enumerate}

Another idea for future work is to add noise and firewall filtering to the studied data.
This would allow the detection of a firewall, to identify different firewalls and to make more robust tests.


\end{document}